\documentclass[aps,prl,preprint,superscriptaddress,amsmath]{revtex4}
\usepackage{longtable}
\usepackage{graphicx}
\usepackage{amsmath,amssymb}
\usepackage{color}
\usepackage{bbm}

\newcommand{\e}{{\rm e}}
\newcommand{\drm}{{\rm d}}

\newcommand{\beq}{\begin{equation}}
\newcommand{\eeq}{\end{equation}}
\newcommand{\bdm}{\begin{displaymath}}
\newcommand{\edm}{\end{displaymath}}

\newcommand{\Cdot}{\!\cdot\!}

\begin{document}

\title{Simulation of underground gravity gradients from stochastic seismic fields}

\author{Jan Harms}
\affiliation{University of Minnesota, 116 Church Street SE, Minneapolis, MN 55455, USA}
\author{Riccardo DeSalvo}
\affiliation{California Institute of Technology, East Bridge, Pasadena, California 91125, USA}
\author{Steven Dorsher}
\affiliation{University of Minnesota, 116 Church Street SE, Minneapolis, MN 55455, USA}
\author{Vuk Mandic}
\affiliation{University of Minnesota, 116 Church Street SE, Minneapolis, MN 55455, USA}


\date{\today}

\begin{abstract}
We present results obtained from a finite-element simulation of seismic displacement fields and of gravity gradients generated by those fields. The displacement field is constructed by a plane wave model with a 3D isotropic stochastic field and a 2D fundamental Rayleigh field. The plane wave model provides an accurate representation of stationary fields from distant sources. Underground gravity gradients are calculated as acceleration of a free test mass inside a cavity. The results are discussed in the context of gravity-gradient noise subtraction in third generation gravitational-wave detectors. Error analysis with respect to the density of the simulated grid leads to a derivation of an improved seismometer placement inside a 3D array which would be used in practice to monitor the seismic field.
\end{abstract}
\pacs{04.80.Nn,09.30.Fn,95.75.Wx}

\maketitle 

\section{Introduction}
Within the next decade, the current generation of gravitational-wave detectors \cite{LSC2009b,LuEA2006,VIR2008,Tat2008} will be upgraded or replaced by a second generation of detectors which are expected to produce a wealth of data from astrophysical events producing gravitational waves (GW) above 10\,Hz and up to a few kHz \cite{Cre2003,LSC2009a}. For the second generation, the limiting noise sources at low frequencies are suspension thermal noise, radiation-pressure noise and seismic noise \cite{RoEA2002,RoEA2004}. A variety of solutions appear plausible to overcome limitations posed by these noise sources to pave the way for a third generation of detectors sensitive below 10\,Hz (www.et-gw.eu). However, gravity-gradient noise (GGN) generated by the stochastic seismic field directly couples to the interferometer's test masses and is predicted to impose a low-frequency barrier for future-generation GW detectors \cite{Sau1984,BeEA1998,HuTh1998,Cre2008}. Theoretically, GGN can be mitigated in many different ways: by intelligent choice of location, by intelligent architecture of buildings and soil, and by subtraction of an estimated GGN contribution from the detector output, or a combination of the above techniques. Since the gravity gradients are linked to the spectrum of seismic displacement, the most obvious strategy is to search for a seismically quiet place. It is confirmed by borehole studies and underground experiments in mines \cite{Bor2002,CarEA1991,Dou1964} that the seismic noise above 1\,Hz and at 1\,km depth is smaller by more than an order of magnitude relative to the surface level. The explanation for this observation is that the surface supports additional surface modes, the Rayleigh and Love modes, and the surface is subject to comparatively violent disturbances from atmospheric pressure fluctuations and human activities which both generate seismic noise that decays rapidly towards greater depths. 

In this paper, results are discussed in the context of GGN subtraction. There is no technology available to directly measure gravity gradients above 1\,Hz with the required precision, i.e.~$\sim5\cdot 10^{-19}(\rm m/s^2)/\sqrt{Hz}$ at 1\,Hz. In fact, any instrument that could measure those gravity gradients would also be sensitive to gravitational waves (this is just true for most interferometer topologies including the Michelson interferometer; specialized interferometers can be insensitive to GWs \cite{TAE2000} or to non-GW displacements \cite{KaCh2004}). Therefore, a straight-forward technique is to monitor the seismic field around the test masses and feed the seismic data into a model for gravity-gradient generation which outputs the data that is to be subtracted from the GW detector output. Theoretically, GGN can be reduced arbitrarily well given ideal instruments and unlimited density of the seismometer array. The key questions that we attempt to answer for a simplified model are to what level one can subtract the GGN in practice and, given a certain subtraction level, how many seismometers are needed and out to which distance from the test masses the seismic field needs to be monitored to reach the required subtraction level. For a typical seismic amplitude spectrum deep underground that is about one order of magnitude weaker than spectra at quiet surface locations, an additional subtraction by two orders of magnitude is required to achieve a sensitivity at 1\,Hz comparable to the near-future sensitivity goals above 10\,Hz. For that reason, we will specify our results to this subtraction level. 

Our simulation, which is based on the assumption that all sources of seismic waves are distant, does not answer all of these questions yet, since the answers are highly dependent on the detector site chosen. For a specific location, one needs to know the spectral densities of the Rayleigh field, the body field and the field generated by local sources, which include scattering centers. Rayleigh scattering, the reflection of seismic waves from small-volume density or Lam\'e-constants perturbations, is known to produce wave attenuation and decrease of the correlation length of the seismic field \cite{Wu1962}. In addition, Rayleigh scattering, as a special form of reflection of seismic waves, exhibits mode conversion between shear and pressure waves. In the context of GGN prediction, it is useless to attempt a simulation of Rayleigh scattering based on a simplified approach. What needs to be done is to construct a detailed model of the local geology including all drifts (horizontal tunnels) and shafts, to run a simulation of the seismic field with that model and to compare the results with measurements. 

This will not just help to determine the quality of a GGN model that just takes fields from distant sources into account, but it will also help to evaluate the quality of needed seismic stations. An abnormally low measured correlation between two seismic instruments would indicate poor quality of the coupling between seismometer and seismic field, and how to improve it. The generally high degree of inhomogeneity of rock or sediments near the Earth surface, including man-made inhomogeneities like foundations of buildings, is suspected to be responsible for the widely reported lack of coherence in surface measurements \cite{Bor2002}. Similar problems occur in underground environments where drilling and blasting creates cracks and density or stress disturbances in the rock close to the fabricated mine workings, which are known to alter the seismic field. It will be one of the future challenges to design seismic stations for ultra-sensitive broadband seismometers avoiding locally generated disturbances of the seismic measurement. 

Another minor simplification is to evaluate the GGN at a single location and to interpret this result as displacement noise in GW detectors. This approach neglects possible correlations between GGN at different test masses. However, above 1\,Hz, correlations are negligible even in hard rock assuming distances of about 10\,km between test masses. This is true for theoretical predictions based on ideal models and especially for real measurements where localized sources, scattering and attenuation further decrease correlations. 

We consider the simplified isotropic, stochastic plane-wave model of the seismic field. It is also assumed that the local geology is uniform in which case the fundamental Rayleigh field is the only surface field. We investigate properties of underground gravity gradients without specifying absolute values for spectral densities. The reader can find estimates in previous literature on gravity gradients \cite{Sau1984,BeEA1998,HuTh1998,Cre2008}. In Sec.~\ref{sec:Theo}, we outline the basics of classical gravity-field perturbations. The model for the surface and body seismic fields used in our simulation are described in Sec.~\ref{sec:Seismic}. Finally, in Sec.~\ref{sec:Res}, we analyze the gravity-gradient production by those fields and present our results.

\section{Theoretical foundations}
\label{sec:Theo}
In the past, the generation of gravity gradients from seismic fields was either linked to the perturbation $\delta\rho$ of a mean density $\rho_0$ of rock as a bulk contribution or to a surface effect where air is replaced by rock due to normal displacement of the surface. We will outline briefly how to calculate gravity gradients from bulk and surfaces and how to combine them into a unified description.

Inside rock, seismic displacement $\vec{\xi}(\vec{r},t)$ generates density perturbations according to
\beq
\delta\rho = -\rho_0\,{\rm div}\vec\xi
\label{eq:Dens}
\eeq
Three different modes contribute to $\vec\xi$. The horizontal (SH) and vertical (SV) shear modes are transverse waves and have equal speed $c_{\rm S}$ in isotropic rock. Their displacement fields are divergence-free and one does not have to include them in gravity-gradient calculations if contributions from surfaces are excluded. They are important only at boundaries between lighter and denser rocks, and above all between soil and air. The third mode is the longitudinal pressure wave P. It has speed $c_{\rm P}>c_{\rm S}$ and is entirely responsible for the density perturbations inside the medium. From Eq.\,(\ref{eq:Dens}), one calculates the gravity gradient at some point $\vec{r}_0$ by
\beq
\delta\vec a_{\rm b}(\vec{r}_0,t) = -G\rho_0\int\drm V \frac{{\rm div}\vec{\xi}(\vec{r},t)}{|\vec r-\vec r_0|^2}\cdot \vec e_r
\label{eq:Div}
\eeq
where $\vec e_r$ is the unit vector $(\vec r-\vec r_0)/|\vec r-\vec r_0|$. We call Eq.\,(\ref{eq:Div}) the divergence or body contribution to the gravity gradient.

In addition to the divergence term, in case of voids or close to the Earth surface, there is a surface contribution to GGN which is related to any displacement normal to the surface:
\beq
\delta\vec a_{\rm s}(\vec{r}_0,t) = G\rho_0\int\drm S \frac{\vec n(\vec{r})\cdot\vec{\xi}(\vec{r},t)}{|\vec r-\vec r_0|^2}\cdot \vec e_r
\label{eq:Surface}
\eeq
This equation is easy to understand. At each point at the surface, a previously empty volume of size $\drm S(\vec n\cdot\vec{\xi})$ is replaced by surface soil of density $\rho_0$, or the displacement is opposite to $\vec n$ so that an empty volume is created. All these contributions are summed up to give the surface gravity gradient. Body and surface contributions as formulated in Eq.\,(\ref{eq:Div}) and Eq.\,(\ref{eq:Surface}) are linear perturbations of the exact theory. While this is obvious for the density perturbations Eq.\,(\ref{eq:Dens}), the surface term neglects higher orders by assuming that the normal vector $\vec n(\vec r)$ does not change with time and is evaluated at the unperturbed location of the surface.

Although the surface and body integrals do not seem to have much in common, it is possible to cast their sum into the simple form
\beq
\begin{split}
\delta\vec{a}(\vec{r}_0,t) &= \delta\vec{a}_{\rm b}(\vec{r}_0,t)+\delta\vec{a}_{\rm s}(\vec{r}_0,t)\\
&= G\rho_0\int\drm
V\dfrac{1}{|\vec r-\vec r_0|^3}\left(\vec{\xi}(\vec{r},t) -3(\vec{e}_r\cdot\vec{\xi}(\vec{r},t))\cdot\vec{e}_r\right)
\end{split}
\label{eq:Dipole}
\eeq
The easiest way to prove this is to perform an integration by parts of Eq.\,(\ref{eq:Div}). The sum formula should be familiar to the reader. It describes a dipole perturbation where in this case the dipole moment is given by the displacement field $\vec\xi$. From a finite-element point of view, one could have guessed this equation right from the beginning. The finite elements are represented by point masses $\rho_0\drm V$ that are displaced by $\vec\xi$. The lowest-order perturbation of a homogenous field due to the displaced point mass must be the dipole field. Since the dipole term comprises body and surface effects, Eq.\,(\ref{eq:Dipole}) is the natural choice to simulate gravity gradients. The remaining problem is to define the displacement field with surface and body fields of different types. This will be described in Sec.~\ref{sec:Seismic}.

\section{Simulation of seismic fields}
\label{sec:Seismic}
Our finite-element (FEM) simulation is purely kinematical, which is much faster than running a fully dynamical simulation with commercial software. Many relevant aspects of the seismic field such as surface reflection and mode conversion, tube waves etc., are well described in many publications and monographs and can be taken care of in the simulation. There is no need to calculate the solution of the displacement field with initial data and boundary conditions. Admittedly, as soon as one attempts to simulate a geologically complex environment, then a dynamical FEM simulation may become necessary. This approach is followed by a group in the Netherlands (led by Jo van den Brand) which attempts to fill in many of the required details coming from rock inhomogeneities.

A kinematical simulation means to use a simple expansion of the field into plane or spherical waves and to propagate them through the grid or along surfaces. We simulate one frequency at a time, which means that the field is characterized by three different wave lengths (i.e. shear, pressure and fundamental Rayleigh waves). The model for the Earth surface is plane and the horizontal directions are denoted by $x,y$, the vertical direction by $z$. The surface level is defined by $z_0=0$, deeper levels have negative $z$ coordinates. As we will show later, it is sufficient to simulate a grid whose size is twice the length of pressure waves in all directions from the test mass for which the gravity gradient is calculated. Surface waves are described by plane or spherical 2D waves exponentially decaying at greater depths (evanescent waves). We also investigated the case when grid spacing becomes similar to or larger than the size of the cavity around the test mass. We ran additional high-density small-volume simulations whenever the cavity size was smaller than the grid spacing to make sure that gravity gradients from density changes close to the test mass or from the cavity surface are estimated correctly. From this study we found that in all cases the corrections from the fine-grid analysis were negligible for the purpose of this paper. We also neglected surface modes of cavities in the case of the test mass being underground \cite{Sti1959,Whi1962}. Though not included in our simulation, the before-mentioned publications clearly indicate that all but one of the additional discrete modes would appear at high frequencies above 100\,Hz. Also, the cavity modes would have to be generated at the cavity surface, and it is reasonable to assume that sources of cavity modes such as pressure fluctuations inside the cavity or machinery directly acting on cavity surfaces can be avoided or eliminated.

Seismic reflections from density changes can be complicated since incident P waves partially convert into SV waves and vice versa (SH waves do not convert). However, reflections from a plane rock-air interface are comparatively simple, and concise analytical solutions exist \cite{Bor1961,Ken2001}. The reflection mechanism can be understood in various ways: energy transport, surface displacement, etc. In our case, the main purpose is to calculate the amplitudes of the reflected modes. The amplified surface displacement due to constructive interference of up-going and down-going waves, which depends on the angle of incidence, results automatically. However, we do not include generation of evanescent P waves from reflected SV waves with nearly horizontal direction of propagation. This parameter regime fills the gap between travelling waves and fully evanescent fields, i.e.~the Rayleigh field. Our approach is to treat evanescent modes separately in terms of an independent Rayleigh field. All in all, investigating the variation of gravity gradients by changing the composition of the seismic field and varying the physical parameters of the rock, we think that our simulation already contains more details than necessary for our aims. Many results are equally valid for Rayleigh waves, as for waves from many randomly localized sources, and as for the stochastic isotropic field.

\subsection{Stochastic isotropic fields}
\label{ssec:Isotropic}
At first, we need to specify what we mean by an isotropic field. Strictly speaking, we do not simulate an isotropic field. Instead, we need to consistently combine the notion of an isotropic field with the boundary conditions at the free surface. The isotropic field in our simulation is formed by plane waves with random directions of propagation, but all emerging from the lower half space. Each wave is reflected from the Earth surface to produce an amplified surface displacement via constructive interference with the reflected wave. In our model, the field's energy of injected waves is uniformly partitioned between the three polarizations SV, SH and P. 

Let us start with a description of the reflection of seismic waves at the free surface. The SH waves are by definition polarized parallel to the surface. No mode conversion takes place and the amplitude of the incident wave is equal to the amplitude of the reflected wave. The wave's phase does not change and surface displacement doubles relative to the amplitude of the SH wave. 

Mode conversion between SV and P makes the reflection problem intractable in general, and still sufficiently complicated for plane, free surfaces. The ansatz to obtain the reflection coefficients is to require vanishing traction at the surface. We refer the reader to the first book of the monograph by B.~Kennett \cite{Ken2001} where this problem is elegantly solved for free surfaces or rock-fluid interfaces. Here, we will just introduce the notation and present the results. The simplest way to parameterize surface phenomena is to introduce the concept of slowness which is the inverse of speed. Given the speed of shear and pressure waves $c_{\rm S},\,c_{\rm P}$ inside the rock, the horizontal slowness $p$, e.g. $p=\sin(\alpha)/c_{\rm P}$, provides a complete parametrization of reflection coefficients, where in this example $\alpha$ is the angle of incidence of a P wave with respect to the vertical direction. Mode conversion is exclusively between waves of the same horizontal slowness. Therefore, it would be more complicated to reformulate the problem in terms of the angle $\alpha$. For example, a P wave incident with angle $\alpha$ gives rise to a P wave reflected at the same angle, but also to a converted mode SV that subtends a smaller angle $\beta(\alpha)<\alpha$. 

For each mode SV and P with common horizontal slowness and phase speeds $c_{\rm S}$ and $c_{\rm P}$, we define the vertical slowness
\beq
q^{\phantom{2}}_{\rm S}(p)=\sqrt{\frac{1}{c_{\rm S}^2}-p^2},\qquad q^{\phantom{2}}_{\rm P}(p)=\sqrt{\frac{1}{c_{\rm P}^2}-p^2}
\label{eq:Slow}
\eeq
Then, the complex reflection coefficients can be cast into the form
\beq
\begin{split}
\rho^{\phantom{2}}_{\rm SS}(p) &= \rho^{\phantom{2}}_{\rm PP}(p)=\dfrac{(2p^2-\frac{1}{c_{\rm S}^2})^2-4p^2q^{\phantom{2}}_{\rm S}q^{\phantom{2}}_{\rm P}}{4p^2q^{\phantom{2}}_{\rm S}q^{\phantom{2}}_{\rm P}+(2p^2-\frac{1}{c_{\rm S}^2})^2}\\
\rho^{\phantom{2}}_{\rm SP}(p) &= \rho^{\phantom{2}}_{\rm PS}(p)=-i\dfrac{4p(2p^2-\frac{1}{c_{\rm S}^2})\sqrt{q^{\phantom{2}}_{\rm S}q^{\phantom{2}}_{\rm P}}}{4p^2q^{\phantom{2}}_{\rm S}q^{\phantom{2}}_{\rm P}+(2p^2-\frac{1}{c_{\rm S}^2})^2}
\end{split}
\label{eq:Refl}
\eeq
These equations can be used for the entire range of horizontal slownesses $0<p<\infty$, but we will restrict to a regime where vertical slownesses of the relevant modes stay real, i.e.\;no evanescent reflections. As mentioned before, the evanescent surface field is treated independently in our simulation. An incident shear wave SV with $1/c_{\rm P}<p<1/c_{\rm S}$ will generate evanescent P waves at reflection ($q_{\rm P}$ becomes imaginary). For $p>1/c_{\rm S}$, SV and P waves are evanescent ($q_{\rm P}$ and $q_{\rm S}$ are imaginary) and combine to form the Rayleigh field. So we will restrict to $p<1/c_{\rm S}$ and ignore the evanescent P mode for $p>1/c_{\rm P}$.
\begin{figure}[ht!]
\center{\includegraphics[width=8cm]{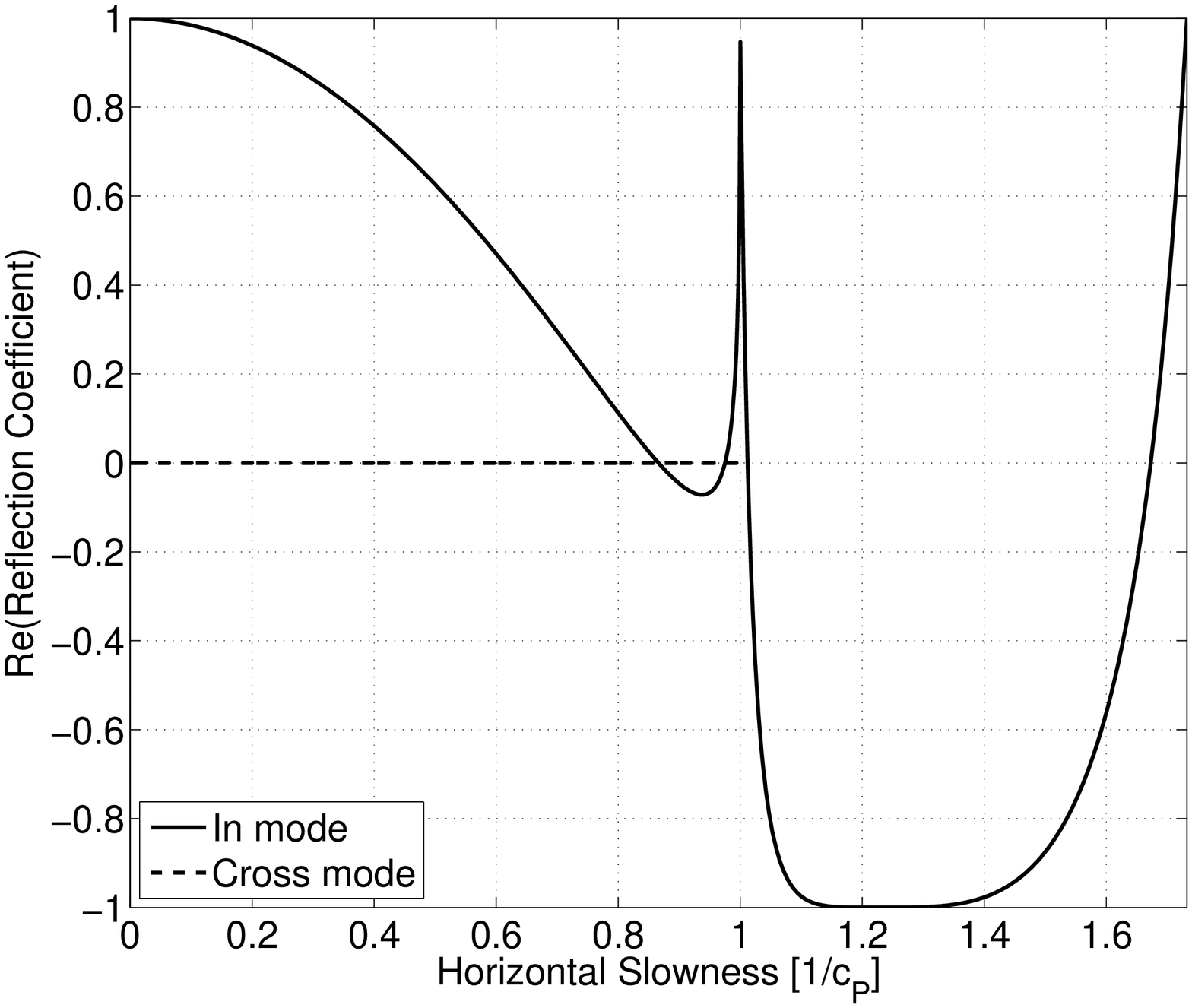}\\
\includegraphics[width=8cm]{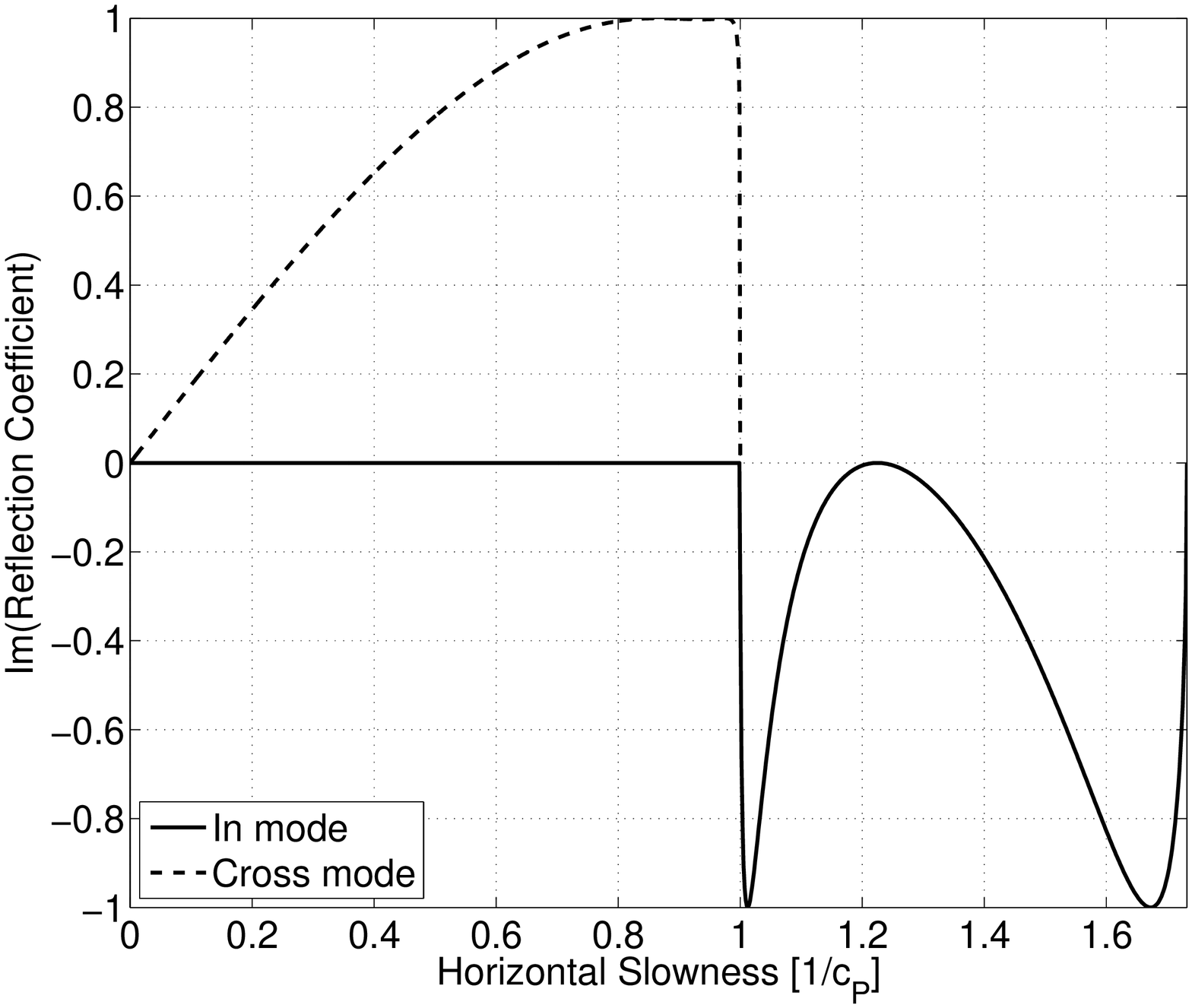}}
\caption{Real and imaginary parts of the reflection coefficients at the free surface. The cross-mode (conversion) coefficient is drawn up to $p=1/c_{\rm P}$ since this value corresponds to horizontal propagation of P waves and also marks the slowness at which reflected SV waves start to be converted into evanescent P waves. Higher slownesses are just needed to describe reflection of SV into SV modes.}
\label{fig:Refl}
\end{figure}
Fig.~\ref{fig:Refl} shows the reflection coefficients for $c_{\rm S}=0.58\cdot c_{\rm P}$ which corresponds to a medium with Poisson ratio $\nu=0.25$ (see Eq.\,(\ref{eq:Speeds})). In that case, the largest angle of incidence of SV waves before the reflected P mode becomes evanescent is $\beta = \arcsin(c_{\rm S}/c_{\rm P})=34.5^\circ$ (reminder: this angle is with respect to normal direction). 

In all examples to follow, the isotropic field is constructed from $N=50$ pressure waves with speed $c_{\rm P}=3\,$km/s and $2\cdot N$ shear waves whose speed is determined by
\beq
c_{\rm S} = c_{\rm P}\cdot\sqrt{\frac{1-2\nu}{2-2\nu}}
\label{eq:Speeds}
\eeq
assuming a Poisson solid ($\nu=0.25,\,c_{\rm S}=0.58\cdot c_{\rm P}$). Directions of propagation are drawn from a uniform distribution over the upper half sphere. Each wave is reflected according to Eq.\,(\ref{eq:Refl}). 

\subsection{Stochastic fields from fundamental Rayleigh waves}
Isotropic media with uniform density do not support Love waves or higher order Rayleigh modes, so that we just need to consider the fundamental Rayleigh mode with our simplified approach. As Rayleigh waves are coherently composed of SV and P waves, even the phase of the displacement field shows vertical dependence because SV and P waves decay at different rates.
\begin{figure}[ht!]
\centerline{\includegraphics[width=8cm]{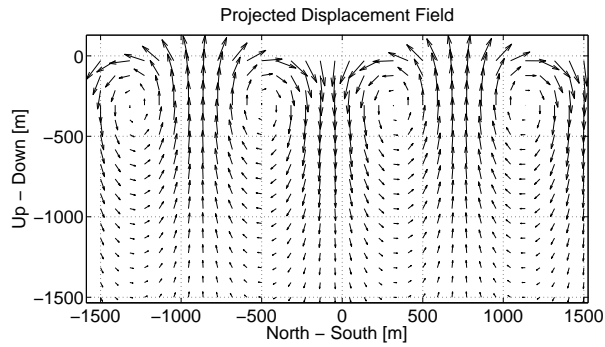}}
\caption{The figure shows a cross-section of the displacement field of a Rayleigh wave. The rate of exponential amplitude decay is different for the pressure and shear part so that the direction of displacement depends on depth.}
\label{fig:Ray}
\end{figure}
Rewriting the equations in \cite{HaNa1998} in terms of the horizontal and vertical slownesses, the horizontal and vertical displacement of the Rayleigh wave read
\beq
\begin{split}
\xi_{\rm hor}&=A\cdot\left(p_{\rm R}\e^{\omega q^\prime_{\rm P}z}-\zeta q^\prime_{\rm S}\cdot\e^{\omega q^\prime_{\rm S}z}\right)\cdot\sin(\phi(\vec r))\\
\xi_{\rm ver}&=A\cdot\left(q^\prime_{\rm P}\e^{\omega q^\prime_{\rm P}z}-\zeta p_{\rm R}\cdot\e^{\omega q^\prime_{\rm S}z}\right)\cdot\cos(\phi(\vec r))
\end{split}
\eeq
with $\phi(\vec r)=\phi_0-\vec k_{\rm R}\cdot\vec r$, $\zeta\equiv\sqrt{q^\prime_{\rm P}/q^\prime_{\rm S}}$, $|\vec k_{\rm R}| = 2\pi f p_{\rm R}$, $\omega = 2\pi f$ and finally $p_{\rm R}=1/c_{\rm R}$. The primed quantities are related to vertical slownesses as defined in Eq.\,(\ref{eq:Slow}) via $q^\prime_{\rm S,P}\equiv-i\cdot q_{\rm S,P}(p_{\rm R})$. The speed $c_{\rm R}$ of the fundamental Rayleigh wave obeys the equation
\beq
\begin{split}
&R\left(\frac{c_{\rm R}}{c_{\rm S}}\right)=0,\\ &R(x)=x^6-8x^4+8x^2\frac{2-\nu}{1-\nu}-\frac{8}{1-\nu}
\end{split}
\eeq
The displacement vector is constructed according to
\beq
\vec\xi(\vec r) = \xi_{\rm hor}(\vec r)\frac{\vec k_{\rm R}}{|\vec k_{\rm R}|}+\xi_{\rm ver}(\vec r)\vec e_z
\eeq
We simulate the total Rayleigh field by summing contributions from 20 plane Rayleigh waves whose directions of propagation are drawn from an isotropic 2D distribution. 

\section{Results}
\label{sec:Res}
In this section, we present our results of a gravity-gradient simulation based on the seismic-field decomposition as specified in the previous section. Unless explicitly stated otherwise, our simulations are based on parameter values as listed in Table \ref{tab:Values}. The Poisson ratio $\nu$ completely determines the ratio of wavelengths of the different wave types P, S and Rayleigh.
\begin{table}[ht!]
\begin{tabular}{|l|r|}
\hline
Parameter & Value \\
\hline\hline
P wavelength $\lambda_{\rm P}$ & 1000\,m\\
\hline
S wavelength $\lambda_{\rm S}$ & 577\,m\\
\hline
Rayleigh wavelength $\lambda_{\rm R}$ & 531\,m\\
\hline
Poisson ratio $\nu$ & 0.25\\
\hline
Mean rock density $\rho_0$ & $2.5\,\rm g/cm^3$\\
\hline
Composition of isotropic field $n_{\rm P}$, $n_{\rm S}$ & 50, 100\\
\hline
Composition of Rayleigh field $n_{\rm R}$ & 20\\
\hline
Grid size & $(2\; {\rm to}\; 6\,\rm km)^3$\\
\hline
Number of grid points & $(100\; {\rm to}\; 150)^3$\\
\hline
\end{tabular}
\caption{The lengths of shear and Rayleigh waves are determined by $\lambda_{\rm P}$ and the Poisson ratio $\nu$. Most results are weakly dependent on the number of plane waves $n_{\rm P}$, $n_{\rm S}$, $n_{\rm R}$ that are used to construct the (isotropic) fields from distant sources.}
\label{tab:Values}
\end{table}

\subsection{A scaling law for the absolute value of GGN}
We present a scaling law that helps to understand the link between seismic fields and gravity-gradient production. It holds under ideal conditions, i.e.\;homogeneous rock, no surfaces, distant sources only. It states that gravity gradients from seismic fields are independent of the seismic wave length (equivalently, seismic speed). By consequence, the gravity-gradient spectrum, measured as amplitude spectral density of acceleration, adopts the frequency dependence of the seismic displacement spectrum. 

\begin{figure}[ht!]
\centerline{\includegraphics[width=8cm]{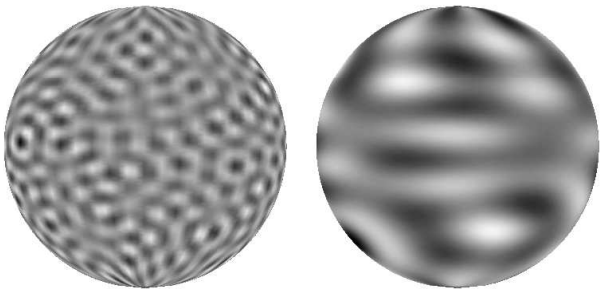}}
\caption{Density perturbations are characterized by a correlation length. If the correlation length is high as represented by the sphere on the right-hand side, then larger volumes of rock are perturbed in a similar way and fewer of those coherent volumes fill the spherical shell. Therefore, integrating density perturbations over shells around the test mass, one would expect that fields with smaller correlation length generate less GGN since perturbations add up incoherently. However, as explained in the text, this conclusion is wrong for most of the fields met in nature.}
\label{fig:Average}
\end{figure}

Figure \ref{fig:Average} illustrates two isotropic density-perturbation fields with different correlation lengths on spheres which have identical radii.
Now, an easy mistake is to compare the cases and to conclude that the field with shorter correlation length produces less GGN at the center of the sphere based on the assumption that having many more cells (a cell comprises rock of similar density perturbation) on the sphere leads to a higher degree of averaging out of gravity gradients. The problem is that one has to consider integration over all spherical shells around the test mass, and we will show that the integral for the class of fields considered here does not depend on wave length.

Let us first introduce the concept of displacement cells. As in Fig.~\ref{fig:Average}, the density-perturbation or displacement field exhibits a certain pattern which consists of distinguishable regions of similar displacement or density perturbation. We call each region a cell and identify it by an index $i$. The volume of each cell is denoted by $V_i$ and its position by $\vec r_i=r_i\vec e_i$. Furthermore, each cell is characterized by a fiducial displacement $\vec \xi_i(t)$ such that the gravity-gradient integral over any subset of cells can be discretized as follows:
\beq
\begin{split}
\delta\vec{a}(\vec 0,t) &= G\rho_0\int\drm
V\dfrac{1}{r^3}\left(\vec{\xi}(\vec{r},t) -3(\vec{e}_r\cdot\vec{\xi}(\vec{r},t))\cdot\vec{e}_r\right) \\
&= G\rho_0\sum\limits_i V_i\dfrac{1}{r_i^3}\left(\vec{\xi}_i(t) -3(\vec{e}_i\cdot\vec{\xi}_i(t))\cdot\vec{e}_i\right)
\end{split}
\eeq
Since the positions of cells are defined by the displacement pattern at a given time, it would be more accurate to understand the cell positions as functions of time, $\vec r_i(t)=r_i(t)\vec e_i(t)$, but, for ease of notation, the time dependence will not be written explicitly. In the following, to simplify the calculation, isotropy and homogeneity of the seismic field is assumed, which entails that the seismic field is characterized by a single correlation length $L$, independent of direction and location. Whereas homogeneity is essential for the argument, a similar calculation can be carried out for arbitrary seismic fields produced by distant sources. From isotropy it follows that the average volume of each cell is simply the cube of the correlation length of the field. Since the correlation length is proportional to the wavelength $\lambda$ of the field, the cell volume is also proportional to the cube of the wavelength: $V_i \propto \lambda^3$. Rescaling the cell positions, $\vec r_i^\prime\equiv \vec r_i/\lambda$, and volumes, $V_i^\prime\equiv V_i/\lambda^3$, one obtains
\beq
\delta\vec{a}(\vec 0,t) = G\rho_0\sum\limits_i V^\prime_i\dfrac{1}{r^{\prime\,3}_i}\left(\vec{\xi}_i(t) -3(\vec{e}_i\cdot\vec{\xi}_i(t))\cdot\vec{e}_i\right)
\label{eq:Cells}
\eeq
The distribution of directions $\vec e_i$ is directly linked to the degree of isotropy of the seismic field, which does not explicitly depend on the lengths of seismic waves, and which does not change under rescaling of the cell positions. For the moment, we also consider the distribution of fiducial displacements $\vec\xi_i(t)$ as independent of the correlation lengths. Since the mean distance of cells is again proportional to the correlation length $L$ and therefore to the wavelength $\lambda$, the rescaled cell positions $\vec r_i^\prime$ are independent of $\lambda$. In total, the gravity-gradient sum Eq.\,(\ref{eq:Cells}) is independent of $\lambda$. 

Depending on the context of this discussion, it can be inappropriate to consider the distribution of displacements $\vec\xi_i$ as being independent of $\lambda$. A simple example is to relate the wavelengths to frequencies through $\lambda f=c$. Since in reality the typical displacement depends on frequency, and therefore in this context on wavelength, Eq.\,(\ref{eq:Cells}) also depends on wavelength. The conclusion from the last paragraph would have to be modified: the spectrum of the gravity gradient $\delta\vec{a}$ adopts the frequency dependence of the spectrum of the displacement field $\vec\xi$. Note that one could add more complexity to the problem by investigating other frequency dependencies like a frequency-dependent degree of isotropy etc. Finally, we want to point out that the requirement of homogeneity also includes the absence of surfaces. A surface introduces an independent length scale $H$ that corresponds to the depth of the test mass, and which makes the sum Eq.\,(\ref{eq:Cells}) dependent on the ratio $H/\lambda$.

\subsection{The isotropic body field}
\label{ssec:IsoGG}
The first question that we address is how large the rock volume is that needs to be monitored with seismic instruments to predict gravity gradients from the isotropic body field with 99\% accuracy for subtraction of GGN. The answer depends on symmetries of the seismic field. For example, if the field was spherically symmetric around the test mass where GGN is calculated, the gravity gradients would vanish. This symmetry property is the reason why P waves have a negligible influence on horizontal GGN near large plane surfaces, since in order to produce surface displacement in the sense of Eq.\,(\ref{eq:Surface}), longitudinal P waves have to propagate almost perpendicularly to the surface, otherwise most of the displacement would be in horizontal directions (one has to keep in mind though that the incident P-wave would be converted partially into an SV wave). Any such wave produces similar displacement over a large area of the surface which is determined by the radius of curvature of the wavefront and the angle of inclination. Integrating gravity gradients over circles on the surface around the test mass, horizontal gravity gradients from opposite sides of the test mass partially cancel. Therefore, horizontal surface GGN as described by Eq.\,(\ref{eq:Surface}) is almost entirely produced by SV waves.

Here, we will answer the first question for a test mass located underground far away from the Earth surface and assuming an isotropic stochastic field as defined in Sec.~\ref{ssec:Isotropic}. As mentioned before, except for a weak contribution from the cavity surface, GGN is determined entirely by the P-wave content of the field.
\begin{figure}[ht!]
\centerline{\includegraphics[width=8cm]{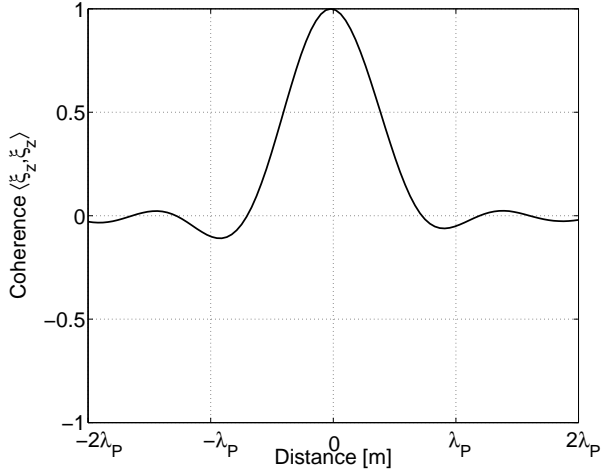}}
\caption{The figure displays the spatial coherence of the P-wave field as a function of distance between the two points of the rock volume forming the correlation pair. We choose those points to be at the same depth (to make the calculation easier) and to correlate the respective $z$-displacements. A similar result would be obtained for the horizontal displacements, although the height of the side maxima is different.}
\label{fig:SpatialCoh}
\end{figure}
As a first step, let us determine the spatial coherence of the P-wave field as a function of distance between two points inside the rock. The spatial coherence as a function of distance provides the length scale that governs the displacement pattern of the isotropic field. It is well known that displacement generated by the isotropic stochastic field is coherent over cells of size $(\lambda/2)^3$ which is in agreement with our results displayed in Fig.~\ref{fig:SpatialCoh}. In other words, the spatial correlation length is tightly linked to the length of the seismic waves, and the question about the size of volume that needs to be monitored by a seismic array can be answered using the wavelength as pertinent length scale. In the following, we will use the P-wave length $\lambda_{\rm P}$ as distance scale. With the help of  Table \ref{tab:Values}, the reader can translate distances into multiples of the lengths of shear or Rayleigh waves. 

\begin{figure}[ht!]
\centerline{\includegraphics[width=8cm]{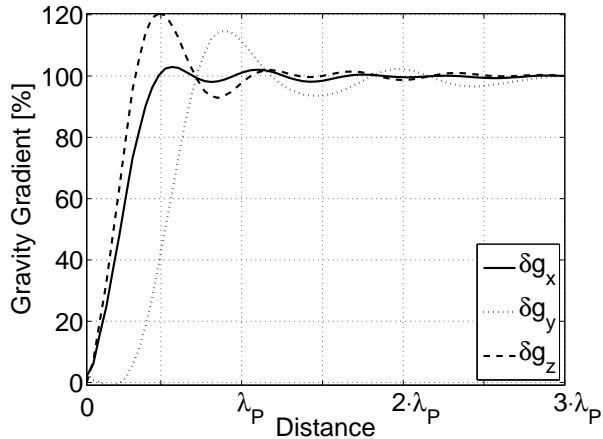}}
\caption{The plot shows the integrated gravity gradients. The curves represent a particular realization of the isotropic field. The test mass is located inside a cavity of radius 2\,m, deep inside the rock (no GGN from Earth surface). More than 90\% of the GGN is contributed from shells within a radius of $\lambda_{\rm P}$. Although the absolute value of the gravity gradient can change from one realization to the next, the curves always converge in the same characteristic way.}
\label{fig:GGIso}
\end{figure}

For a specific realization of the isotropic field, the gravity gradient integrated over spherical mass shells around the test mass as a function of the shell's radius is plotted in Fig.~\ref{fig:GGIso}. We see that the gravity gradient converges to its final value within a sphere of radius $3\Cdot\lambda_{\rm P}$. Residual contributions beyond $3\Cdot\lambda_{\rm P}$ are well below the 1\% level. For the isotropic field, it is possible to construct a 99\% accurate GGN model by monitoring a volume of rock that is considerably smaller than $(3\Cdot\lambda_{\rm P})^3$. The convergence curves of the gravity gradient as shown in Fig.~\ref{fig:GGIso} always assume the same oscillatory shape. Instead of increasing the distance to high values to obtain the limit of the gradient, one can also average the oscillation of the curve within some distance interval closer to the test mass. For example, if the seismic field was monitored and accurately modelled out to a distance $1.5\Cdot\lambda_{\rm P}$, then one could take the corresponding gradient curve and calculate the average of the values between $\lambda_{\rm P}$ and $1.5\Cdot\lambda_{\rm P}$. Doing this we found that the average value approximates the limit of the curve with an accuracy better than 99\%.
These results also demonstrate that a passive suppression of the GGN by means of a large cavity around the test mass would be inefficient. Even if we assume that the cavity has a radius of 100\,m, only a small fraction of the gravity gradient would be eliminated. This conclusion stays true for any kind of rock that is hard enough to provide the stability required to support such a cavity. 

\subsection{Contributions from the Rayleigh field to underground gravity gradients}
\begin{figure}[ht!]
\centerline{\includegraphics[width=8cm]{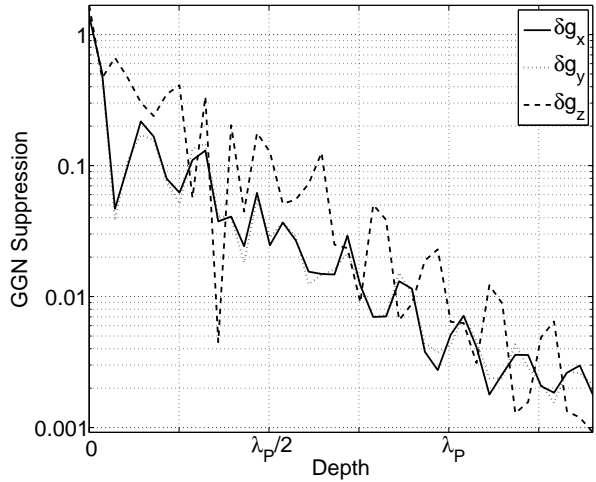}}
\caption{Gravity gradients fall exponentially with increasing depth. Each point in this graph is an average over 50 realizations of the Rayleigh field, each composed of 20 plane Rayleigh waves. Assuming that the surface GGN is difficult to model from seismic data and that the surface seismicity is a factor of 10 -- 30 stronger than the ambient body field, one needs to bring the test masses to underground levels where surface GGN is suppressed by a factor of 1000 -- 3000 to reach the 1\% level. However, by means of a surface array of seismometers it should be possible to model part of the GGN from Rayleigh waves and to relax the requirements.}
\label{fig:SurfaceGG}
\end{figure}

The contribution of Rayleigh waves to gravity gradients at underground levels is difficult to model because the Earth's surface is often far from flat, and local sources of surface waves are numerous and lead to complex seismic fields. The ideal case is if gravity gradients from surface waves could be neglected altogether. As a first step, we look at the gravity gradients from 2D isotropic surface Rayleigh fields shown in Fig.~\ref{fig:SurfaceGG}. Each value is an average over 50 different realizations of the field. Gravity gradients fall exponentially with increasing depth and reach a suppression of 1000 at a depth of about $1.5\cdot\lambda_{\rm P}$. 
Assuming that amplitude spectral densities of surface displacement are about a factor 10 to 30 higher than underground, the suppression of surface GGN would have to be between 1000 to 3000 in order to be a negligible contribution to the total GGN (down to the 1\% level). In most hard rocks near the surface, P-wave speed is about 3 -- 6\,km/s in which case a surface GGN suppression of 3000 at 3\,Hz is reached at a depth of about 1.5 -- 3\,km. It is unlikely that a gravitational-wave detector will be built at 3\,km depth and therefore one would have to look for a place that has no more than average surface seismicity, and/or deploy a surface array of seismometers in addition to the underground array to improve modelling of gravity gradients from surface waves. Especially for the lowest frequencies, surface features become less important and fewer sensors would be needed. 

We should mention that surface contributions to GGN from an isotropic Rayleigh field is subject to cancelling effects when integrating over the entire surface. As was pointed out to us by Jo van den Brand, localized sources at the surface produce gravity gradients that can be significant at underground levels. One of the challenges of future experimental studies will be to investigate the nature and strength of local surface sources and whether a sufficiently simple model can be constructed to calculate their gravity-gradient fields.

\subsection{Coarse-grid errors and seismometer placement}
\label{ssec:Array}
In this subsection, we will investigate how errors build up in gravity-gradient integrals Eq.\,(\ref{eq:Dipole}) depending on the density of the simulation grid $n_{\rm gr}=\drm N/\drm V$. The idea is to start with a uniform grid of maximal possible density, and then to compare the corresponding gravity gradient with gravity gradients obtained from grids with fewer grid points with uniform or non-uniform grid spacing. We will interpret the displacement of the coarse grid as ideal, noise-free seismic measurements whereas the fine grid represents the real seismic field. Gravity gradients are compared by producing curves as shown in Fig.~\ref{fig:GGIso} for the coarse and fine grids, though leaving out the normalization to 100\%, and to evaluate the error from making the grid coarser. Since part of the calculation involves interpolation of coarse grids back to the original grid density which is a memory-consuming operation in Matlab, a smaller number of grid points, $70^3$, is used, and to preserve a sufficiently high spatial resolution close to the test mass, we choose a comparatively small grid volume $(3\,\rm km)^3$. 

\begin{figure}[ht!]
\centerline{\includegraphics[width=8cm]{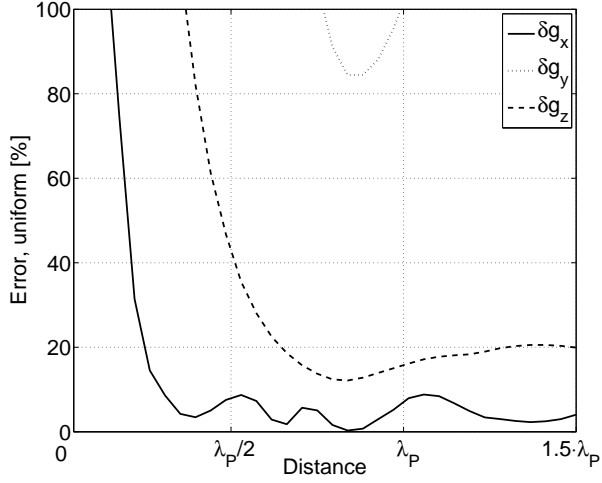}}
\caption{The figure shows the error of the gravity-gradient integral as a function of integration distance. In this case, the original uniform fine grid is compared with a uniform coarse grid. Integration errors of the coarse grid can accumulate to high values as in $\delta g_y$ for this particular realization of the seismic field.}
\label{fig:ErrUni}
\end{figure}

First, we consider the reduction to a coarse, uniform grid with $15^3$ grid points. Calculating the gravity gradients $\delta g_{\rm fi},\,\delta g_{\rm co}$ from the fine and coarse grid for a particular realization of the stochastic isotropic field --- again neglecting the Rayleigh surface field --- the error as percentage
\beq
e=100\cdot\left|\dfrac{\delta g_{\rm fi}-\delta g_{\rm co}}{\delta g_{\rm fi}}\right|
\label{eq:Error}
\eeq
is plotted in Fig.\,\ref{fig:ErrUni} for each of the components $\delta g_{\rm x},\delta g_{\rm y},\delta g_{\rm z}$ as a function of integration distance. More clearly, gravity gradients are integrated up to some variable distance from the test mass, and for each integral the value of the error curve is determined by Eq.\,(\ref{eq:Error}).
As one can see, reduction to a uniform coarse grid leads to very large errors which can easily exceed 100\%. Details of the error curve depend strongly on the realization of the displacement field, but all error curves for the uniform grid have in common that a large error builds up at close distance to the test mass which is sometimes compensated to some extent by contributions from greater distances (as for $\delta g_{\rm x},\delta g_{\rm z}$ in Fig.~\ref{fig:ErrUni}), but which can also stay at a very high level like the $\delta g_{\rm y}$ curve. 

If we regard the position of grid points in the coarse grid as locations of seismometers in a uniform seismometer array and the displacement of those grid points as the seismometer data, then this result means that one cannot use the data of a uniform array to accurately predict the gravity gradient, at least not without applying a more sophisticated model of the seismic field. Also, the number of seismometers, $N=15^3=3375$, is very high. The question is whether $N$ can be reduced, simultaneously decreasing the error.

\begin{figure}[ht!]
\centerline{\includegraphics[width=8cm]{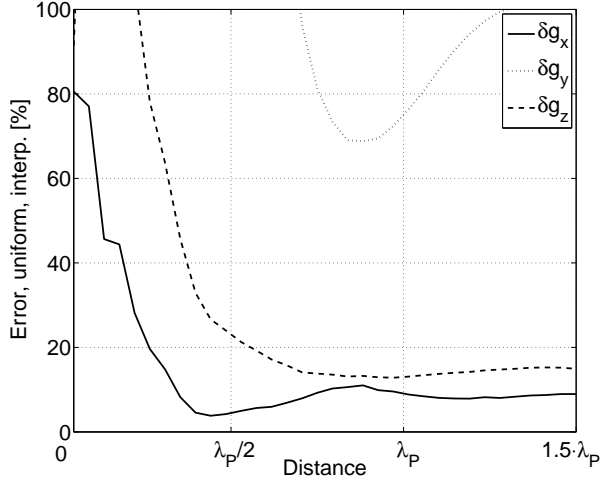}}
\caption{When the displacement field represented by the uniform coarse grid is interpolated back to the original grid density, the error of the gravity-gradient integral decreases. In the majority of realizations, like this one, the improvement is not significant.}
\label{fig:ErrUniInterp}
\end{figure}
\begin{figure}[ht!]
\centerline{\includegraphics[width=8cm]{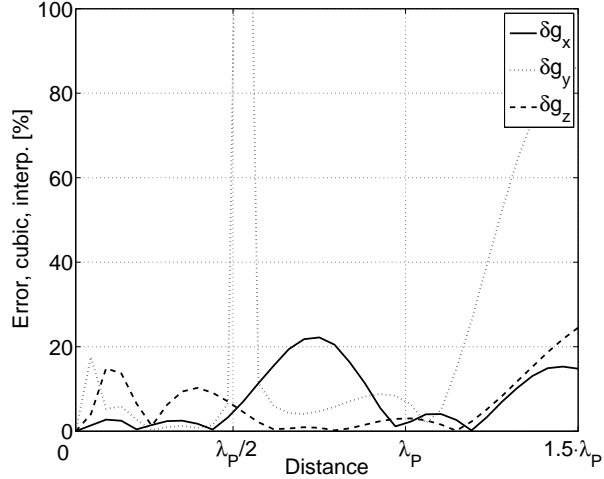}}
\caption{The cubed selection flattens the error curves compared to the linear selection, but does not lead to a decreased error of the gravity-gradient integral. The reason is that grid density is too small starting at distances greater than about $\lambda_{\rm P}/2$. Although the error is significantly better than in the uniform case at intermediate distances (near $\lambda_{\rm P}$), convergence of the error curves towards increasing distance is very poor. The cubed selection does not produce satisfying results, at least not without applying more sophisticated models (yet to be developed) of the seismic field (instead of simple interpolation between grid points).}
\label{fig:ErrCubedInterp}
\end{figure}
\begin{figure}[ht!]
\centerline{\includegraphics[width=8cm]{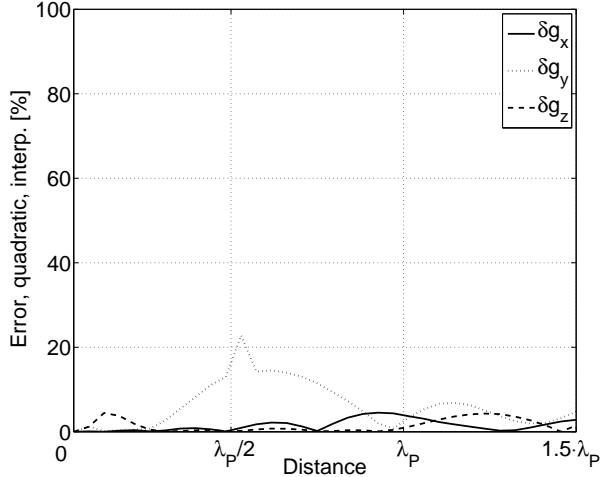}}
\caption{The squared selection produces very good results. The error curves are flat and errors do not build up at greater distances despite the decreased grid density. These properties were observed in all realizations of the isotropic field.}
\label{fig:ErrSquareInterp}
\end{figure}

One idea is to make use of the fact that the displacement field varies smoothly between grid points and to interpolate the coarse-grid displacement field back to the original grid density before calculating the gravity-gradient integral. The trend is that the error decreases by only a small amount. For particular realizations of the seismic field we observed a much better improvement of the error by interpolation, but according to Fig.~\ref{fig:ErrUniInterp}, it is not guaranteed that interpolation decreases the error to a sufficiently small level. Now, the previous error curves and the $1/r^3$ dependence of the integrand Eq.\,(\ref{eq:Dipole}) suggest that increasing the density of grid points near the test mass may flatten the error curves close to the test mass and remove part of the error. In this paper, we investigate two alternative selections of grid points which lead to non-uniform coarse grids: the squared and the cubed selection.
\begin{figure}[ht!]
\centerline{\includegraphics[width=8cm]{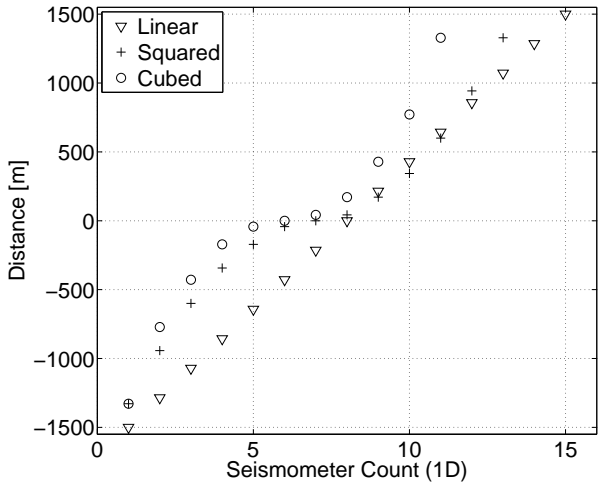}}
\caption{A linear selection of grid points of the fine grid leads to the uniform, coarse grid which is investigated at the beginning of this section. Instead, one can pick more grid points close to the test mass and fewer grid points at greater distances. For example, in what we call the cubed selection, the distance between neighboring grid points is increased to about 500\,m close to the boundary of the grid whereas the distance of grid points close to the test mass is about 40\,m. Grid points in the uniform coarse grid have constant distance of about 220\,m to their neighbors.}
\label{fig:Select}
\end{figure}
Figure \ref{fig:Select} shows how grid points are selected from the fine grid to obtain the respective coarse grid. The squared selection corresponds to $n_{\rm gr}\propto r^{-1}$, and the cubed selection to $n_{\rm gr}\propto r^{-2}$ which means that grid density is increased close to the test mass and decreased near the grid boundary. As a result, the squared coarse grid contains $13^3=2197$ grid points (selection is along Cartesian axes, and not spherically symmetric), and the cubed selection contains $11^3=1331$ grid points. In both cases, the displacement field is interpolated back to a uniform fine grid with $70^3$ grid points before calculating the gravity gradient. The error curves in Fig.~\ref{fig:ErrCubedInterp} for the cubed selection  prove that increasing the density of grid points close to the test mass eliminates all of the low-distance error seen in the uniform coarse grid. 
However, even if the grid density of the cubed selection is increased globally by a constant factor, a recurring problem is that errors build up at greater distances due to the very low grid density. Therefore, we introduce the squared selection, and finally the error curves have an acceptable shape as demonstrated in Fig.~\ref{fig:ErrSquareInterp}.
Although it is tempting to conclude from Fig.~\ref{fig:ErrSquareInterp} that modelling of gravity gradients from isotropic seismic fields (or any other field generated by distant sources) is feasible, and even relatively simple, by clever sensor positioning and interpolation of the seismic data, one should not forget that these results are obtained from a rather large number of grid points / seismometers. The hardware cost of a high number of seismometers is secondary, but the question is whether a suitable borehole system around the test mass can be constructed to house the seismic stations. This kilometer-size, dense array of seismometers around each vertex of a 10\,km triangle would represent a seismic telescope of unprecedented capabilities. In any case, at this point the results should be considered solely relevant for simulation purposes and maybe to give an idea of the problems that will be encountered in future theoretical and experimental studies. We are convinced that the modelling methods will have to be develop to a much higher level than presented in this paper.

\section{Conclusion}
\label{sec:Discuss}
Mitigation of gravity-gradient noise is one of the major challenges in the development of future-generation GW detectors with sensitivity below 10\,Hz. In order to extend below 10\,Hz strain sensitivities which are comparable to those of surface detectors at 100\,Hz, subtraction of GGN from the interferometer data will be necessary. Gravity-gradient noise cannot be directly measured by any other instrument than a GW detector itself. Therefore, the GGN prediction must be based on knowledge of its sources, which is the seismic field (and atmospheric density fluctuations if a surface location of the detector is considered). Sources of the seismic field are either distant, in which case a plane-wave model of the seismic field can be constructed, or local, which requires more detailed understanding of the source and its radiation characteristics. In this paper, we constructed seismic fields from a plane-wave model and investigated the associated underground gravity gradients. We found that the isotropic, stochastic Rayleigh field produces small gravity gradients at depths comparable to the length of pressure waves. Depending on the level of seismic activity at the surface and of the frequency considered, these gravity gradients can become totally negligible, i.e. smaller than 1\% of gravity gradients from the ambient body-wave fields. More specifically, we showed that at a depth corresponding to 3/2 of the length of a pressure wave, gravity gradients from the Rayleigh field are suppressed by a factor 1000. Neglecting the surface field, a key question, which we addressed using our simplified model, was how many seismometers need to be deployed to obtain sufficient information about the seismic field for a sufficiently accurate gravity-gradient model, and how large the monitored rock volume needs to be. The scaling law derived in this paper shows that for a certain (relative) subtraction goal the volume that needs to be monitored scales with the cube of the wavelength, and for each given frequency the number of seismometers that need to be deployed is independent of the length or frequency of the seismic waves. For a very simple interpolation method of seismic data, we found that good gravity-gradient modelling results can be obtained if the density of seismic instruments in the array is proportional to the inverse of the distance to the test mass with a total number of seismometers of about 1500. Theoretically, this number is much higher than the information content in the simulated seismic field, so we expect that analyzing the seismic data with a plane-wave model of the seismic field as opposed to simple interpolation will achieve similar performance with a considerably smaller number of seismometers. For the future, our work needs to be developed further by investigating the fields from local sources, including Rayleigh scattering and reflections from fault planes, and by implementation of a seismic model when it comes to analyzing seismic data, not just to simulate the seismic field. Investigation and comparison of local sources in surface and underground environments is the key component to understand the advantages and disadvantages of building underground detectors. 

\section{Acknowledgments}
We gratefully acknowledge the support of LIGO and thank Albrecht R\"udiger for his thorough scrutiny of this paper. 

\raggedright
\bibliography{c:/MyStuff/references}

\end{document}